# A FORMALIZATION OF TOP-DOWN UNNESTING

THOMAS NEUMANN

Abstract. When writing SQL queries, it is often convenient to use correlated subqueries. However, for the database engine, these correlated queries are very difficult to evaluate efficiently. The query optimizer will therefore try to eliminate the correlations, a process referred to as *unnesting*.

Recent work has introduced a single-pass top-down algorithm for unnesting arbitrary SQL queries. That work did not include a formal proof of correctness, though. In this work we provide the missing formalization by formally defining the operator semantics and proving that the unnesting algorithm is correct.

## 1. Introduction

Subqueries can be used in most places within a SQL query, and subqueries can be nested within each other. Furthermore, subqueries can be *correlated*, which means that the subquery references attributes provided by the outer query. We can see an example with two nested correlated subqueries below:

```
SELECT *
FROM customer
WHERE c_mktsegment='AUTOMOBILE' AND
  (SELECT COUNT(*)
   FROM orders
   WHERE o_custkey=c_custkey AND
     (SELECT SUM(l_extendedprice)
      FROM lineitem
      WHERE l_orderkey=o_orderkey)>300000)>5
```

For the runtime system this query is very challenging, as a naive evaluation would execute the correlated subqueries for each tuple of the corresponding outer queries, which leads to very poor performance on even moderately sized data sets.

Usually the query optimizer will try to remove the correlations, a process referred to as *unnesting*, but many database systems can only unnest some particular instances of correlated queries. A fully general approach for unnesting arbitrary queries was presented in [1], but the approach leads to suboptimal results when correlated subqueries are nested within each other, as query fragments are unnested repeatedly. Recent work [2] fixed this limitation by providing a single-pass, top-down variant of this unnesting strategy that can handle these nested queries. No formal proof of correctness was given, though.

In this paper we provide a formalization of the unnesting approach from [1] and top-down unnesting from [2]. It is meant to be read as complementary material, i.e., it proves the correctness of the approach but does not introduce the algorithm itself.

The text consists of proofs that build upon each other, thus it has to be written in bottom-up fashion, proving increasingly larger pieces until we can show that the whole algorithm is correct. We will first introduce the notation used within the paper in Section 2 and then we define several relational operators and show some basic properties in Section 3. Afterwards we prove that the individual unnesting steps are correct in Section 4. Using that, we can show that the whole algorithm correctly unnests arbitrary queries in Section 5.

## 2. Notation

A *tuple* $t$ is an unordered mapping from attribute names to values. We define $A(t)$ as the set of names of all attributes contained in $t$. We can access the value of a single attribute $a$ contained in $A(t)$ by writing $t.a$. We can construct a tuple $t$ with a single attribute $a$ and associated value $x$ by writing $t := [a : x]$. Given two tuples $t_1$ and $t_2$ such that $A(t_1) \cap A(t_2) = \emptyset$, we can concatenate both into a new tuple $t$ by writing $t := t_1 \circ t_2$. Given a tuple $t$, we can restrict $t$ to a subset of attributes $A' \subseteq A(t)$ by writing

$$t_{|A'} := \circ_{a \in A'} [a : t.a] \tag{1}$$

A *relation* $R$ with associated schema $A(R)$ is a multi-set of tuples such that for every $t \in R$ the condition $A(t) = A(R)$ holds. Note that in the following we will implicitly assume that all operations are on multi-sets. That is, our sets can contain duplicates unless explicitly noted otherwise.

When reasoning about multi-sets it is often useful to operate using the *characteristic function*. For a given multi-set $A$, we define the characteristic function $m_A(x)$ as

$$m_A(x) := \begin{cases} n \text{ if } A \text{ contains } x \quad n \text{ times} \\ 0 \text{ otherwise} \end{cases} \tag{2}$$

Note that the characteristic function uniquely identifies a multi-set, that is, we can define a multi-set by specifying a characteristic function.

For a given scalar expression $e$, we write $F(e)$ to compute the set of *free variables*, i.e., the attributes that are referenced by $e$. We generalize this concept to relational expressions by writing $F(R)$ to compute the free variables of the whole relational expression $R$, i.e., the attributes that are referenced within $R$ but that are not defined by an operator in $R$. Note that we cannot evaluate a relational expression with free variables. For example, given two relations $R_1$ and $R_2$, with $A(R_1) = \{x\}$ and $A(R_2) = \{y\}$, and relational expression $T := \sigma_{x=y}(R_2)$, we have $F(T) = \{x\}$, and thus $T$ cannot be evaluated in itself. But for a larger expression $T' := R_1 \Join T$, we have $F(T') = \emptyset$, because $x$ became bound by the (dependent) join, and $T'$ can be evaluated.

## 3. Relational operators

Using this notation, we can now define the relational operators. For the standard set operations, given two relations $R_1$ and $R_2$ with $A(R_1) = A(R_2)$, we define the set operations by specifying their characteristic functions:

$$\begin{aligned} m_{R_1 \cup R_2}(x) &:= m_{R_1}(x) + m_{R_2}(x) \\ m_{R_1 \cap R_2}(x) &:= \min\Big(m_{R_1}(x), m_{R_2}(x)\Big) \\ m_{R_1 \setminus R_2}(x) &:= \max\Big(m_{R_1}(x) - m_{R_2}(x), 0\Big) \end{aligned} \tag{3}$$

For a given relation $R$, we define the *duplicate eliminating projection* $\Pi$ on a given set of attributes $A'$ with $A' \subseteq A(R)$ as

$$\Pi_{A'}(R) := \Big\{x \mid \exists y \in R : x = y_{|A'}\Big\} \tag{4}$$

Note that the result is a set and not a multi-set. Using that definition, we define the *duplicate preserving projection* $\Pi^+$ on a given set of attributes $A'$ with $A' \subseteq A(R)$ as

$$m_{\Pi^+_{A'}(R)}(x) := \sum_{x' \in \Pi_{A(R)}(R)} m_R(x') \cdot 1_{x'_{|A'}=x}. \tag{5}$$

Note that this definition uses the *indicator function* $1_p := \begin{cases} 1 \text{ if } p \\ 0 \text{ otherwise} \end{cases}$ to keep the definition shorter.

Given a relation $R$, and a filter predicate $p$ with $F(p) \subseteq A(R)$, we define the *selection* $\sigma_p$ as

$$\begin{aligned} m_{\sigma_{p(R)}}(x) &:= \begin{cases} m_R(x) & \text{if } p(x) \\ 0 & \text{otherwise} \end{cases} \\ &= m_R(x) \cdot 1_{p(x)} \end{aligned} \tag{6}$$

Given a relation $R$, and a map expression $a : f$ with $a \notin A(R)$ and $F(f) \subseteq A(R)$, we define a *map* $\chi_{a:f}$ as

$$m_{\chi_{a:f}(R)}(x) := \sum_{x' \in \Pi_{A(R)}(R)} m_R(x') \cdot 1_{x = x' \circ [a:f(x')]}. \tag{7}$$

Note that this definition implicitly adds the attribute $a$, computed as $f(x')$, to the resulting relation.

With that we can now define a rename operator in terms of projection and map. Given a relation $R$, and two attribute names $a$ and $a'$ with $a \notin A(R)$ and $a' \in A(R)$, we define a *rename* $\rho_{a:a'}$ as

$$\rho_{a:a'}(R) := \Pi^+_{(A(R) \setminus \{a'\}) \cup \{a\}}(\chi_{a:a'}(R)). \tag{8}$$

Given two relations $R_1$ and $R_2$ with $A(R_1) \cap A(R_2) = \emptyset$, we can define the *cross product* $\times$ by specifying that $A(R_1 \times R_2) = A(R_1) \cup A(R_2)$ and defining

$$m_{R_1 \times R_2}(x) := m_{R_1}\left(x_{|A(R_1)}\right) \cdot m_{R_2}\left(x_{|A(R_2)}\right) \tag{9}$$

We can now define the *inner join* $\bowtie$ of two relations $R_1$ and $R_2$ with $A(R_1) \cap A(R_2) = \emptyset$ with a join condition $p$ as

$$R_1 \bowtie_p R_2 := \sigma_p(R_1 \times R_2). \tag{10}$$

For the next definition we need a helper definition that we call *bind*. The bind operator $\text{bind}(R, t)$ takes a relational expression $R$ and a tuple $t$ and evaluates $R$ by making the attributes of $t$ globally available. In particular, the attributes from $A(t)$ are not considered free variables during the evaluation, as their definition is known. As bind only substitutes the attributes of $t$ as constants, it commutes with all relational operators, e.g., $\text{bind}(R_1 \cup R_2, t) = \text{bind}(R_1, t) \cup \text{bind}(R_2, t)$ and $\text{bind}\big(\sigma_p(R), t\big) = \sigma_p(\text{bind}(R, t))$.

With that we can now define the *dependent join* $\mathbin{\blacktriangleright\!\triangleleft}$ of a relation $R_1$ and a relational expression $R_2$ with $A(R_1) \cap A(R_2) = \emptyset$ and $F(R_2) \subseteq A(R_1)$ (ignoring globally bound attributes) on a join condition $p$, analogous to the regular join and regular cross product:

$$\begin{aligned} m_{R_1 \mathbin{\blacktriangleright\!\triangleleft} R_2}(x) &:= m_{R_1}\left(x_{|A(R_1)}\right) \cdot m_{\text{bind}\left(R_2, x_{|F(R_2)}\right)}\left(x_{|A(R_2)}\right) \\ m_{R_1 \mathbin{\blacktriangleright\!\triangleleft}_p R_2}(x) &:= m_{R_1 \mathbin{\blacktriangleright\!\triangleleft} R_2}(x) \cdot 1_{p(x)} \end{aligned} \tag{11}$$

We can show that a dependent join is equivalent to a regular join if the right hand side does not depend on the left hand side:

**Lemma 3.1.** *Given two relations $R_1$ and $R_2$ with $A(R_1) \cap A(R_2) = \emptyset$ and $F(R_2) = \emptyset$, and a predicate $p$ with $F(p) \subseteq A(R_1) \cup A(R_2)$. Then $R_1 \bowtie_p R_2 \equiv R_1 \mathbin{\blacktriangleright\!\triangleleft}_p R_2$.*

*Proof.* We will show the equivalence by showing that $m_{R_1 \bowtie_p R_2} = m_{R_1 \mathbin{\blacktriangleright\!\triangleleft}_p R_2}$. From Equation 10 and Equation 11 we can deduce that it is sufficient to show that $m_{R_1 \times R_2} = m_{R_1 \mathbin{\blacktriangleright\!\triangleleft} R_2}$, as both characteristic functions are 0 if $p(x)$ is false. We show that by the following expansion

$$
\begin{aligned}
m_{R_1 \mathbin{\blacktriangleright\!\triangleleft} R_2} &= m_{R_1}\left(x_{|A(R_1)}\right) \cdot m_{\text{bind}\left(R_2, x_{|F(R_2)}\right)}\left(x_{|A(R_2)}\right) && \text{(Equation 11)} \\
&= m_{R_1}\left(x_{|A(R_1)}\right) \cdot m_{R_2}\left(x_{|A(R_2)}\right) && \text{(as } F(R_2) = \emptyset\text{)} \\
&= m_{R_1 \times R_2} && \text{(Equation 9)}
\end{aligned}
$$

□

As a syntactical shorthand we allow for declaring a join condition as *natural*, for example like this: $R_1 \bowtie_{p \wedge \text{ natural}} R_2$. This computes the common attributes $C := A(R_1) \cap A(R_2)$, creates a unique symbol $a'_i$ for each $a_i \in C$, renames the attributes $a_i \in C$ in $R_2$ into the corresponding $a'_i$, adds the condition $a_i = a'_i$ to the join condition for each $a_i$, and eliminates the $a'_i$ in the result by using $\Pi^+_{A(R_1) \cup A(R_2)}$. Note that in the context of NULL values this equality condition $a_i = a'_i$ considers two NULL values to be equal, i.e., NULL is simply treated as an additional value in the domain.

A natural join can be expressed easily using the characteristic functions:

**Lemma 3.2.** *Given two relations $R_1$ and $R_2$. Then*

$$
m_{R_1 \bowtie_{\text{natural}} R_2}(x) = m_{R_1}\left(x_{|A(R_1)}\right) \cdot m_{R_2}\left(x_{|A(R_2)}\right). \tag{12}
$$

*Proof.* We define the set of common attributes as $C := A(R_1) \cap A(R_2)$. The claim holds trivially if $C = \emptyset$. In the following we assume $C$ is non-empty. Then, expanding the *natural* shorthand and using that the conditions $a_i = a'_i$ functionally determine the renamed attributes $a'_i$ (so that $\Pi^+$ preserves multiplicities),

$$
m_{R_1 \bowtie_{\text{natural}} R_2}(x) = \begin{cases} m_{R_1}\left(x_{|A(R_1)}\right) \cdot m_{R_2}\left(x_{|A(R_2)}\right) \text{ if natural } (x) \text{ holds} \\ 0 \text{ otherwise} \end{cases}
$$

By definition the *natural* condition introduces new symbols $a'_i$ for each attribute $a_{i'}$ in $C$, renames all attributes $a_i$ in $R_2$ into $a'_i$ and then checks that $x.a_i = x.a'_i$ for each attribute. Thus

$$
m_{R_1 \bowtie_{\text{natural}} R_2}(x) = \begin{cases} m_{R_1}\left(x_1 := x_{|A(R_1)}\right) \cdot m_{R_2}\left(x_2 := x_{|A(R_2)}\right) \text{ if } \forall_{a_i \in C} x_1.a_i = x_2.a_i \\ 0 \text{ otherwise} \end{cases}.
$$

Note that $x_1$ and $x_2$ are projections of $x$, thus $x.a_i = x_1.a_i$ and $x.a_i = x_2.a_i$ holds for every $a_i$ in $C$. By transitivity, this implies that $x_1.a_i = x_2.a_i$ for every $a_i$ in $C$. Therefore the if condition is always true, and the claim follows.

□

To support the SQL `GROUP BY` construct we define a *group by* operator $\Gamma_{A;a:f}(R)$ where $A \subseteq A(R)$ and $a \notin A(R)$ and $f$ is an aggregation function on $R$ as

$$
\Gamma_{A;a:f}(R) := \left\{ x \circ [a : f(R')] \mid x \in \Pi_A(R) \wedge R' = \sigma_{\forall_{a_i \in A} a_i = x.a_i}(R) \right\} \tag{13}
$$

Note that this operator produces a set and not a multi-set. As for scalar expressions, we require that an aggregation function only observes the attributes it references. That is, $f$ references a fixed set of attributes $F(f) \subseteq A(R)$, and its value is fully determined by these attributes, i.e., $f(S) = f\left(\Pi^+_{F(f)}(S)\right)$ for every relation $S$ with $F(f) \subseteq A(S)$. In particular, extending the input of $f$ with additional attributes does not change the result of the aggregation. This holds for all standard scalar aggregates like, e.g., `SUM`. For aggregates like, e.g., `json_agg`, the system has to explicitly fix the referenced columns.

For additional SQL constructs we also define some derived operators that can be expressed using the already defined ones. Their unnesting steps follow from the underlying base operators, but we will briefly show the resulting equivalences in Section 4.

First we have the *semi join* $\ltimes$ and *anti join* $\triangleright$ defined on two relations $R_1$ and $R_2$ with $A(R_1) \cap A(R_2) = \emptyset$ and predicate $p$:

$$\begin{aligned} R_1 \ltimes_p R_2 &\coloneqq R_1 \cap \Pi^+_{A(R_1)}\big(R_1 \bowtie_p R_2\big) \\ R_1 \triangleright_p R_2 &\coloneqq R_1 \setminus \big(R_1 \ltimes_p R_2\big) \end{aligned} \tag{14}$$

To define outer joins we introduce a helper function *nullpad* that gets a set of attribute names $A$ and sets all attributes in $A$ that are not already defined in the input to NULL. Note that this is syntactic sugar for a sequence of maps:

$$\text{nullpad}_A(R) \coloneqq \begin{cases} R & \text{if } A \subseteq A(R) \\ \text{nullpad}_A\Big(\chi_{\left(\arg\min_{a_i \in (A \setminus A(R))} i\right):\text{NULL}}(R)\Big) & \text{otherwise} \end{cases} \tag{15}$$

Then the (left) *outer join* $⟕$ can be defined as:

$$R_1 \mathbin{⟕}_p R_2 \coloneqq \big(R_1 \bowtie_p R_2\big) \cup \Big(\text{nullpad}_{A(R_2)}\big(R_1 \triangleright_p R_2\big)\Big) \tag{16}$$

## 4. Proving the Unnesting Steps

The unnesting strategy crucially depends on the fact that it is possible to decompose a dependent join into a regular join and a dependent join of the domain of the free variables. Thus, we will prove that decomposition first.

**Theorem 4.1.** *Given a relation $R_1$, a relational expression $R_2$ with $A(R_1) \cap A(R_2) = \emptyset$ and $F(R_2) \subseteq A(R_1)$ and $A^D \coloneqq F(R_2)$, a predicate $p$ with $F(p) \subseteq A(R_1) \cup A(R_2)$, and a duplicate free relation $D$ with $A(D) = A^D$ and $D \supseteq \Pi_{A^D}(R_1)$, the following holds:*

$$R_1 \mathbin{⧑}_p R_2 \equiv R_1 \bowtie_{p \wedge \text{ natural}} (D \mathbin{⧑} R_2) \tag{17}$$

*Proof.* From the definition of $\bowtie$ and $⧑$ we see that $R_1 \bowtie_p R_2 \equiv \sigma_{p(R_1 \times R_2)}$ and $R_1 \mathbin{⧑}_p R_2 \equiv \sigma_{p(R_1 \mathbin{⧑} R_2)}$. Thus, without loss of generality, we can assume that $p$ is always true. If not we move it into a separate $\sigma$ above the respective joins (this is valid because $F(p) \subseteq A(R_1) \cup A(R_2)$, so $p$ does not reference the attributes eliminated by the *natural* projection) and only have to show the equivalence of the joins. This simplifies the claim to

$$R_1 \mathbin{⧑} R_2 \equiv R_1 \bowtie_{\text{natural}} (D \mathbin{⧑} R_2).$$

By expanding the definitions and using Lemma 3.2 we can write the claim as

$$\begin{aligned} & m_{R_1}\Big(x_{|A(R_1)}\Big) \cdot m_{\text{bind}\left(R_2, x_{|F(R_2)}\right)}\Big(x_{|A(R_2)}\Big) \\ = {} & m_{R_1}\Big(x_{|A(R_1)}\Big) \cdot \Big(m_D\Big(x_{|A^D}\Big) \cdot m_{\text{bind}\left(R_2, x_{|F(R_2)}\right)}\Big(x_{|A(R_2)}\Big)\Big) \end{aligned}$$

By comparing both sides we see that the claim follows when $m_{R_1}\Big(x_{|A(R_1)}\Big) > 0$ implies $m_D\Big(x_{|A^D}\Big) = 1$. We will prove that implication by contradiction.

Assume that there exists an $x$ such that $m_{R_1}\left(x_{|A(R_1)}\right) > 0$ but $m_D\left(x_{|A^D}\right) \neq 1$. We know that $D$ is duplicate free, thus $m_D(x')$ is either 0 or 1 for every possible $x'$. Thus $m_D\left(x_{|A^D}\right)$ must be 0. We know that $D$ is a superset of $\Pi_{A^D}(R_1)$, thus $\forall_x m_D(x') \geq m_{\Pi_{A^D}(R_1)}(x')$. Therefore $m_{\Pi_{A^D}(R_1)}\left(x_{|A^D}\right) = 0$. But that is a contradiction to the assumption that $m_{R_1}\left(x_{|A(R_1)}\right) > 0$, as $A^D \subseteq A(R_1)$. The claim follows. □

Note that in Theorem 4.1 we allowed $D$ to be a superset of $\Pi_{A^D}(R_1)$. Thus the unnesting algorithm might produce more tuples than strictly necessary, as long as $D$ stays duplicate free and the values from $\Pi_{A^D}(R_1)$ are preserved. We will make use of this fact when using "perfect" unnesting, i.e., renaming columns instead of introducing a join. For that unnesting step we will use a helper lemma:

**Lemma 4.2.** *Given a duplicate free relation $D$ with $A(D) = \{d\}$, a relation $R$ with $A(R) \cap A(D) = \emptyset \wedge a \in A(R)$, and two relational expressions $T_1 \coloneqq \sigma_{d=a}(D \times R)$ and $T_2 \coloneqq \sigma_{d=a}(\chi_{d:a}(R))$. Then $T_2 \supseteq T_1$, and $m_{T_1}(x) > 0 \Rightarrow m_{T_2}(x) = m_{T_1}(x)$.*

*Proof.* Note that $m_{T_1}(x) > 0 \Rightarrow m_{T_2}(x) = m_{T_1}(x)$ implies that $T_2 \supseteq T_1$, thus we only have to prove the former.

We prove this by contradiction. Assume that there exists an $x$ such that $m_{T_1}(x) > 0$ but $m_{T_2}(x) \neq m_{T_1}(x)$. Due to $m_{T_1}(x) > 0$, we know that $m_D\left(x_{|A(D)}\right) = 1$ (as the only other option would be $m_D\left(x_{|A(D)}\right) = 0$, and that would imply $m_{T_1}(x) = 0$). Then we can expand both side of the inequality as follows

$$
\begin{aligned}
m_{\sigma_{d=a}(D\times R)}(x) &= m_D\left(x_{|A(D)}\right) \cdot m_R\left(x_{|A(R)}\right) \cdot 1_{x.d=x.a} \\
&= m_R\left(x_{|A(R)}\right) \cdot 1_{x.d=x.a} \qquad \left(\text{as } m_D\left(x_{|A(D)}\right) = 1\right) \\
m_{\sigma_{d=a}(\chi_{d:a}(R))}(x) &= \sum_{x' \in \Pi_{A(R)}(R)} m_R(x') \cdot 1_{x_{|A(R)\cup\{d\}} = x' \circ [d:x'.a]} \cdot 1_{x.d=x.a} \\
&= \sum_{x' \in \Pi_{A(R)}(R)} m_R(x') \cdot 1_{x_{|A(R)} = x'} \cdot 1_{x.d=x'.a} \cdot 1_{x.d=x.a} \\
&= \sum_{x' \in \Pi_{A(R)}(R)} m_R(x') \cdot 1_{x_{|A(R)} = x'} \cdot 1_{x.d=x.a} \\
&= m_R\left(x_{|A(R)}\right) \cdot 1_{x.d=x.a}
\end{aligned}
$$

This is a contradiction to the assumption that $m_{T_2}(x) \neq m_{T_1}(x)$, and the claim follows. □

Next we will prove that the dependent join with $D$ can be moved down the relational algebra expression, until we reach a state where the right hand side of the join does not depend on the left hand side. At this point we can eliminate the dependent join according to Lemma 3.1. We will show this for each operator individually, starting with projections.

**Lemma 4.3.** *Given a duplicate free relation $D$ and a relational expression $T \coloneqq \Pi_A(R)$ with $A(D) \cap A(R) = \emptyset$ and $F(T) \subseteq A(D)$. Then $D \Join (\Pi_A(R)) \equiv \Pi_{A\cup A(D)}(D \Join R)$.*

*Proof.* We know that $D$ is duplicate free and that $T$ evaluates to a set. Thus, $\forall_x m_D(x) \leq 1$, and $\forall_{x\in D, x'} m_{\Pi_A\left(\text{bind}\left(R, x_{|F(R)}\right)\right)}(x') \leq 1$. Accordingly, $D \Join T$ is a set, too. We can therefore write down the expression in set notation, which simplifies the transformation:

$$
\begin{aligned}
D \Join (\Pi_A(R)) &= \Big\{ l \circ r \mid l \in D \land r \in \Pi_A\Big(\mathrm{bind}\Big(R, l_{|F(R)}\Big)\Big) \Big\} \\
&= \Big\{ l \circ r \mid l \in D \land r \in \Big\{ x | \exists y \in \mathrm{bind}\Big(R, l_{|F(R)}\Big) : x = y_{|A} \Big\} \Big\} \\
&= \Big\{ l \circ r_{|A} \mid l \in D \land r \in \mathrm{bind}\Big(R, l_{|F(R)}\Big) \Big\} \\
&= \Big\{ x | \exists y \in \Big\{ l \circ r \mid l \in D \land r \in \mathrm{bind}\Big(R, l_{|F(R)}\Big) \Big\} : x = y_{|A \cup A(D)} \Big\} \\
&= \Pi_{A \cup A(D)}(D \Join R)
\end{aligned}
$$

□

**Lemma 4.4.** *Given a duplicate free relation $D$ and a relational expression $T \coloneqq \Pi^+_A(R)$ with $A(D) \cap A(R) = \emptyset$ and $F(T) \subseteq A(D)$. Then $D \Join (\Pi^+_A(R)) \equiv \Pi^+_{A \cup A(D)}(D \Join R)$.*

*Proof.*
We can show this by using Equation 5.

$$
\begin{aligned}
m_{D \Join (\Pi^+_A(R))}(x) &= m_D\Big(x_{|A(D)}\Big) \cdot m_{\Pi^+_A\left(\mathrm{bind}\left(R, x_{|F(R)}\right)\right)}\Big(x_{|A}\Big) \\
&= m_D\Big(x_{|A(D)}\Big) \cdot \sum_{x' \in \left(\Pi_{A(R)}\left(\mathrm{bind}\left(R, x_{|F(R)}\right)\right)\right)} m_{\mathrm{bind}\left(R, x_{|F(R)}\right)}(x') \cdot 1_{x'_{|A} = x_{|A}} \\
&= \sum_{x' \in \left(\Pi_{A(D) \cup A(R)}(D \Join R)\right)} m_{D \Join R}(x') \cdot 1_{x'_{|A(D) \cup A} = x} \\
&= m_{\Pi^+_{A \cup A(D)}(D \Join R)}(x)
\end{aligned}
$$

□

For the set operations the dependent join has to be replicated on both sides of the set operation.

**Lemma 4.5.** *Given a duplicate free relation $D$ and a relational expression $T \coloneqq R_1 \cup R_2$ with $A(D) \cap A(T) = \emptyset$ and $F(T) \subseteq A(D)$. Then $D \Join (R_1 \cup R_2) \equiv (D \Join R_1) \cup (D \Join R_2)$.*

*Proof.* We will show this using Equation 3 and Equation 11:

$$
\begin{aligned}
m_{D \Join (R_1 \cup R_2)} &= m_D\Big(x_{|A(D)}\Big) \cdot \Big( m_{\mathrm{bind}\left(R_1, x_{|F(R_1)}\right)}\Big(x_{|A(T)}\Big) + m_{\mathrm{bind}\left(R_2, x_{|F(R_2)}\right)}\Big(x_{|A(T)}\Big) \Big) \\
&= m_D\Big(x_{|A(D)}\Big) \cdot m_{\mathrm{bind}\left(R_1, x_{|F(R_1)}\right)}\Big(x_{|A(T)}\Big) + m_D\Big(x_{|A(D)}\Big) \cdot m_{\mathrm{bind}\left(R_2, x_{|F(R_2)}\right)}\Big(x_{|A(T)}\Big) \\
&= m_{D \Join R_1}(x) + m_{D \Join R_2}(x) \\
&= m_{(D \Join R_1) \cup (D \Join R_2)}(x)
\end{aligned}
$$

□

**Lemma 4.6.** *Given a duplicate free relation $D$ and a relational expression $T \coloneqq R_1 \cap R_2$ with $A(D) \cap A(T) = \emptyset$ and $F(T) \subseteq A(D)$. Then $D \Join (R_1 \cap R_2) \equiv (D \Join R_1) \cap (D \Join R_2)$.*

*Proof.* We will show this using Equation 3 and Equation 11 and using the fact that $m_D(x) \geq 0$:

$$
\begin{aligned}
m_{D \bowtie (R_1 \cap R_2)} &= m_D\left(x_{|A(D)}\right) \cdot \min\left(m_{\text{bind}\left(R_1, x_{|F(R_1)}\right)}\left(x_{|A(T)}\right), m_{\text{bind}\left(R_2, x_{|F(R_2)}\right)}\left(x_{|A(T)}\right)\right) \\
&= \min\left(m_D\left(x_{|A(D)}\right) \cdot m_{\text{bind}\left(R_1, x_{|F(R_1)}\right)}\left(x_{|A(T)}\right), m_D\left(x_{|A(D)}\right) \cdot m_{\text{bind}\left(R_2, x_{|F(R_2)}\right)}\left(x_{|A(T)}\right)\right) \\
&= \min\left(m_{D \bowtie R_1}(x), m_{D \bowtie R_2}(x)\right) \\
&= m_{(D \bowtie R_1) \cap (D \bowtie R_2)}(x)
\end{aligned}
$$

□

**Lemma 4.7.** *Given a duplicate free relation $D$ and a relational expression $T \coloneqq R_1 \setminus R_2$ with $A(D) \cap A(T) = \emptyset$ and $F(T) \subseteq A(D)$. Then $D \bowtie (R_1 \setminus R_2) \equiv (D \bowtie R_1) \setminus (D \bowtie R_2)$.*

*Proof.* We will show this using Equation 3 and Equation 11:

$$
\begin{aligned}
m_{D \bowtie (R_1 \setminus R_2)} &= m_D\left(x_{|A(D)}\right) \cdot \max\left(m_{\text{bind}\left(R_1, x_{|F(R_1)}\right)}\left(x_{|A(T)}\right) - m_{\text{bind}\left(R_2, x_{|F(R_2)}\right)}\left(x_{|A(T)}\right), 0\right) \\
&= \max\left(m_D\left(x_{|A(D)}\right) \cdot m_{\text{bind}\left(R_1, x_{|F(R_1)}\right)}\left(x_{|A(T)}\right) - m_D\left(x_{|A(D)}\right) \cdot m_{\text{bind}\left(R_2, x_{|F(R_2)}\right)}\left(x_{|A(T)}\right), 0\right) \\
&= \max\left(m_{D \bowtie R_1}(x) - m_{D \bowtie R_2}(x), 0\right) \\
&= m_{(D \bowtie R_1) \setminus (D \bowtie R_2)}(x)
\end{aligned}
$$

□

Next we will show selection and map.

**Lemma 4.8.** *Given a duplicate free relation $D$ and a relational expression $T \coloneqq \sigma_p(R)$ with $A(D) \cap A(T) = \emptyset$ and $F(T) \subseteq A(D)$. Then $D \bowtie \left(\sigma_p(R)\right) \equiv \sigma_p(D \bowtie R)$.*

*Proof.* On a pure syntax level this transformation is valid because the attributes from $A(D)$ are available both on the right hand of the $\bowtie$ (due to the implied *bind*) and in the result of the $\bowtie$. We now have to prove the equivalence of the sets:

$$
\begin{aligned}
m_{D \bowtie \sigma_p(R)}(x) &= m_D\left(x_{|A(D)}\right) \cdot \left(m_{\text{bind}\left(R, x_{|F(R)}\right)}\left(x_{|A(R)}\right) \cdot 1_{p(x)}\right) \\
&= \left(m_D\left(x_{|A(D)}\right) \cdot m_{\text{bind}\left(R, x_{|F(R)}\right)}\left(x_{|A(R)}\right)\right) \cdot 1_{p(x)} \\
&= m_{\sigma_p(D \bowtie R)}(x)
\end{aligned}
$$

□

**Lemma 4.9.** *Given a duplicate free relation $D$ and a relational expression $T \coloneqq \chi_{a:f}(R)$ with $A(D) \cap A(T) = \emptyset$, $a \notin A(R) \cup A(D)$, and $F(T) \subseteq A(D)$. Then $D \bowtie \left(\chi_{a:f}(R)\right) \equiv \chi_{a:f}(D \bowtie R)$.*

*Proof.* This transformation is syntactically valid with the same arguments as in Lemma 4.8. We can prove equivalence by expanding the operators:

$$
\begin{aligned}
m_{D \mathbin{\blacktriangleright\!\triangleleft} \chi_{a:f}(R)}(x) = \quad & m_D\big(x_{|A(D)}\big) \cdot \sum_{x' \in \Pi_{A(R)}\left(\mathrm{bind}\left(R, x_{|F(R)}\right)\right)} m_{\mathrm{bind}\left(R, x_{|F(R)}\right)}(x') \cdot \\
& \qquad 1_{x_{|A(T)} = x' \circ \left[a : \mathrm{bind}\left(f(x'), x_{|F(T)}\right)\right]} \\
= \quad & m_D\big(x_{|A(D)}\big) \cdot \sum_{x' \in \Pi_{A(D) \cup A(R)}(D \mathbin{\blacktriangleright\!\triangleleft} R)} m_{\mathrm{bind}\left(R, x_{|F(R)}\right)}\big(x'_{|A(R)}\big) \cdot \\
& \qquad 1_{x_{|A(D) \cup A(T)} = x' \circ [a : f(x')]} \\
= \quad & \sum_{x' \in \Pi_{A(D) \cup A(R)}(D \mathbin{\blacktriangleright\!\triangleleft} R)} m_D\big(x_{|A(D)}\big) \cdot m_{\mathrm{bind}\left(R, x_{|F(R)}\right)}\big(x'_{|A(R)}\big) \cdot \\
& \qquad 1_{x_{|A(D) \cup A(T)} = x' \circ [a : f(x')]} \\
= \quad & m_{\chi_{a:f}(D \mathbin{\blacktriangleright\!\triangleleft} R)}(x)
\end{aligned}
$$

Note that the $x'$ are bound to unique values of the input in all cases. The transformation makes use of the fact that the expression $x_{|A(D) \cup A(T)} = x' \circ [a : f(x')]$ will only be true for $x'$ values that are identical to $x$ for all $a_i \in A(D)$, thus the first two sums have the same non-zero terms.

□

For cross products and joins we first consider the case that only one side has free variables and then handle the general case. We always assume that our sub-expressions are syntactically valid, e.g., when we have a sub-expression $T \coloneqq R_1 \times R_2$ we implicitly require that $A(R_1) \cap A(R_2) = \emptyset$.

**Lemma 4.10.** *Given a duplicate free relation $D$ and a relational expression $T \coloneqq R_1 \times R_2$ with $A(D) \cap A(T) = \emptyset$ and $F(T) \subseteq A(D)$. Then $F(R_2) = \emptyset \Rightarrow D \mathbin{\blacktriangleright\!\triangleleft} (R_1 \times R_2) \equiv (D \mathbin{\blacktriangleright\!\triangleleft} R_1) \times R_2$, and $F(R_1) = \emptyset \Rightarrow D \mathbin{\blacktriangleright\!\triangleleft} (R_1 \times R_2) \equiv R_1 \times (D \mathbin{\blacktriangleright\!\triangleleft} R_2)$*

*Proof.* Case 1: $F(R_2) = \emptyset$. Then the transformation into $(D \mathbin{\blacktriangleright\!\triangleleft} R_1) \times R_2$ is syntactically valid because $R_2$ does not reference attributes from $A(D)$. Expanding the definitions we can show the equivalence:

$$
\begin{aligned}
m_{D \mathbin{\blacktriangleright\!\triangleleft} (R_1 \times R_2)}(x) &= m_D\big(x_{|A(D)}\big) \cdot \Big(m_{\mathrm{bind}\left(R_1, x_{|F(R_1)}\right)}\big(x_{|A(R_1)}\big) \cdot m_{R_2}\big(x_{|A(R_2)}\big)\Big) \\
&= \Big(m_D\big(x_{|A(D)}\big) \cdot m_{\mathrm{bind}\left(R_1, x_{|F(R_1)}\right)}\big(x_{|A(R_1)}\big)\Big) \cdot m_{R_2}\big(x_{|A(R_2)}\big) \\
&= m_{(D \mathbin{\blacktriangleright\!\triangleleft} R_1) \times R_2}(x)
\end{aligned}
$$

Case 2: $F(R_1) = \emptyset$. Follows by symmetry from Case 1. □

**Lemma 4.11.** *Given a duplicate free relation $D$ and a relational expression $T \coloneqq R_1 \bowtie_p R_2$ with $A(D) \cap A(T) = \emptyset$ and $F(T) \subseteq A(D)$. Then $F(R_2) = \emptyset \Rightarrow D \mathbin{\blacktriangleright\!\triangleleft} \big(R_1 \bowtie_p R_2\big) \equiv (D \mathbin{\blacktriangleright\!\triangleleft} R_1) \bowtie_p R_2$, and $F(R_1) = \emptyset \Rightarrow D \mathbin{\blacktriangleright\!\triangleleft} \big(R_1 \bowtie_p R_2\big) \equiv R_1 \bowtie_p (D \mathbin{\blacktriangleright\!\triangleleft} R_2)$*

*Proof.* Follows from Equation 10, Lemma 4.8 and Lemma 4.10. □

If both sides of a join or cross product depend on $D$ we cannot simply push down the dependent join. Instead, we replicate it on both sides and add a natural join condition.

**Lemma 4.12.** *Given a duplicate free relation $D$ and a relational expression $T \coloneqq R_1 \times R_2$ with $A(D) \cap A(T) = \emptyset$ and $F(T) \subseteq A(D)$. Then $D \mathbin{\blacktriangleright\!\triangleleft} (R_1 \times R_2) \equiv (D \mathbin{\blacktriangleright\!\triangleleft} R_1) \bowtie_{\mathrm{natural}} (D \mathbin{\blacktriangleright\!\triangleleft} R_2)$*

*Proof.* We show the equivalence by transforming the right side of the equivalence into the left side. During the transformation we make use of the fact that $D$ is duplicate free, which means that $m_D(x)$ is either 0 or 1. As a consequence, $\forall_x : m_D(x)^2 = m_D(x)$. This allows us to write

$$
\begin{aligned}
m_{(D \mathbin{\blacktriangleright\!\triangleleft} R_1) \bowtie_{\text{natural}} (D \mathbin{\blacktriangleright\!\triangleleft} R_2)}(x) &= m_{D \mathbin{\blacktriangleright\!\triangleleft} R_1}\left(x_{|A(D) \cup A(R_1)}\right) \cdot m_{D \mathbin{\blacktriangleright\!\triangleleft} R_2}\left(x_{|A(D) \cup A(R_2)}\right) \qquad \text{(Lemma 3.2)} \\
&= m_D\left(x_{|A(D)}\right) \cdot m_{\text{bind}\left(R_1, x_{|F(R_1)}\right)}\left(x_{|A(R_1)}\right) \cdot m_D\left(x_{|A(D)}\right) \cdot m_{\text{bind}\left(R_2, x_{|F(R_2)}\right)}\left(x_{|A(R_2)}\right) \\
&= m_D\left(x_{|A(D)}\right) \cdot m_{\text{bind}\left(R_1, x_{|F(R_1)}\right)}\left(x_{|A(R_1)}\right) \cdot m_{\text{bind}\left(R_2, x_{|F(R_2)}\right)}\left(x_{|A(R_2)}\right) \\
&= m_{D \mathbin{\blacktriangleright\!\triangleleft} (R_1 \times R_2)}(x)
\end{aligned}
$$

□

**Lemma 4.13.** *Given a duplicate free relation $D$ and a relational expression $T := R_1 \bowtie_p R_2$ with $A(D) \cap A(T) = \emptyset$ and $F(T) \subseteq A(D)$. Then $D \mathbin{\blacktriangleright\!\triangleleft} \left(R_1 \bowtie_p R_2\right) \equiv (D \mathbin{\blacktriangleright\!\triangleleft} R_1) \bowtie_{p \wedge \text{ natural}} (D \mathbin{\blacktriangleright\!\triangleleft} R_2)$*

*Proof.* Follows from Equation 10, Lemma 4.8 and Lemma 4.12. □

For group by operators we have to update the group by attributes, as the aggregation has to be performed for each binding separately.

**Lemma 4.14.** *Given a duplicate free relation $D$ and a relational expression $T := \Gamma_{A;a:f} R$ with $A(D) \cap A(R) = \emptyset$, $a \notin A(D)$, and $F(T) \subseteq A(D)$. Then $D \mathbin{\blacktriangleright\!\triangleleft} \left(\Gamma_{A;a:f}(R)\right) \equiv \Gamma_{A \cup A(D);a:f}(D \mathbin{\blacktriangleright\!\triangleleft} R)$*

*Proof.* We know that both $D$ and the result of a $\Gamma$ operator are sets, thus the result of $D \mathbin{\blacktriangleright\!\triangleleft} \left(\Gamma_{A;a:f}(R)\right)$ is a set, too, and we can formulate the transformation using set notation. In the third step the set $R'$ that is computed for a given $x$ is not literally the set $R'$ computed for the pair $l := x_{|A(D)}, r := x_{|A}$ in the step before: the constraints on the $\sigma$ operator put the two into a multiplicity-preserving bijection $s \mapsto l \circ s$, and as $D$ is duplicate free the multiplicities agree. The set in the third step merely carries the additional attributes $A(D)$. The aggregation $f$ yields the same value on both, because $f$ is determined by $\Pi^+_{F(f)}$ of its input (see the definition of $\Gamma$) and $F(f) \subseteq A(R)$ is disjoint from $A(D)$.

$$
\begin{aligned}
D \mathbin{\blacktriangleright\!\triangleleft} \left(\Gamma_{A;a:f}(R)\right) &= D \mathbin{\blacktriangleright\!\triangleleft} \left\{ x \circ [a : f(R')] \mid x \in \Pi_A(R) \wedge R' = \sigma_{\forall_{a_i \in A} a_i = x.a_i}(R) \right\} \\
&= \left\{ l \circ r \circ \left[a : \text{bind}\left(f(R'), l_{|F(T)}\right)\right] \mid l \in D \wedge r \in \Pi_A\left(\text{bind}\left(R, l_{|F(T)}\right)\right) \wedge R' = \sigma_{\forall_{a_i \in A} a_i = r.a_i}\left(\text{bind}\left(R, l_{|F(T)}\right)\right) \right\} \\
&= \left\{ x \circ [a : f(R')] \mid x \in \Pi_{A \cup A(D)}(D \mathbin{\blacktriangleright\!\triangleleft} R) \wedge R' = \sigma_{\forall_{a_i \in (A \cup A(D))} a_i = x.a_i}(D \mathbin{\blacktriangleright\!\triangleleft} R) \right\} \\
&= \Gamma_{A \cup A(D);a:f}(D \mathbin{\blacktriangleright\!\triangleleft} R)
\end{aligned}
$$

□

For the derived join operators we can show the equivalences by expanding the definitions and using the already proven equivalences of the underlying operators:

**Lemma 4.15.** *Given a duplicate free relation $D$ and a relational expression $T := R_1 \ltimes_p R_2$ with $A(D) \cap (A(R_1) \cup A(R_2)) = \emptyset$, and $F(T) \subseteq A(D)$. Then $D \mathbin{\blacktriangleright\!\triangleleft} \left(R_1 \ltimes_p R_2\right) \equiv (D \mathbin{\blacktriangleright\!\triangleleft} R_1) \ltimes_{p \wedge \text{ natural}} (D \mathbin{\blacktriangleright\!\triangleleft} R_2)$. For the special case that $F(R_2) = \emptyset$ we have $D \mathbin{\blacktriangleright\!\triangleleft} \left(R_1 \ltimes_p R_2\right) \equiv (D \mathbin{\blacktriangleright\!\triangleleft} R_1) \ltimes_p R_2$*

*Proof.* We can show this using Equation 14, Lemma 4.6, Lemma 4.4, and Lemma 4.13:

$$\begin{aligned}
D \Join \big(R_1 \ltimes_p R_2\big) &= D \Join \Big(R_1 \cap \Pi^+_{A(R_1)}\big(R_1 \bowtie_p R_2\big)\Big) \\
&= (D \Join R_1) \cap \Big(D \Join \Pi^+_{A(R_1)(R_1 \bowtie_p R_2)}\Big) \\
&= (D \Join R_1) \cap \Pi^+_{A(D) \cup A(R_1)}\big((D \Join R_1) \bowtie_{p \wedge \text{ natural}} (D \Join R_2)\big) \\
&= (D \Join R_1) \ltimes_{p \wedge \text{ natural}} (D \Join R_2)
\end{aligned}$$

If $F(R_2) = \emptyset$, we can use a simpler transformation using Lemma 4.11:

$$\begin{aligned}
D \Join \big(R_1 \ltimes_p R_2\big) &= D \Join \Big(R_1 \cap \Pi^+_{A(R_1)}\big(R_1 \bowtie_p R_2\big)\Big) \\
&= (D \Join R_1) \cap \Big(D \Join \Pi^+_{A(R_1)(R_1 \bowtie_p R_2)}\Big) \\
&= (D \Join R_1) \cap \Pi^+_{A(D) \cup A(R_1)}\big((D \Join R_1) \bowtie_p R_2\big) \\
&= (D \Join R_1) \ltimes_p R_2
\end{aligned}$$

□

**Lemma 4.16.** *Given a duplicate free relation $D$ and a relational expression $T := R_1 \triangleright_p R_2$ with $A(D) \cap (A(R_1) \cup A(R_2)) = \emptyset$, and $F(T) \subseteq A(D)$. Then $D \Join \big(R_1 \triangleright_p R_2\big) \equiv (D \Join R_1) \triangleright_{p \wedge \text{ natural}} (D \Join R_2)$. For the special case that $F(R_2) = \emptyset$ we have $D \Join \big(R_1 \triangleright_p R_2\big) \equiv (D \Join R_1) \triangleright_p R_2$*

*Proof.* We can show this using Equation 14, Lemma 4.7, and Lemma 4.15:

$$\begin{aligned}
D \Join (R_1 \triangleright R_2) &= D \Join \big(R_1 \setminus \big(R_1 \ltimes_p R_2\big)\big) \\
&= (D \Join R_1) \setminus \big((D \Join R_1) \ltimes_{p \wedge \text{ natural}} (D \Join R_2)\big) \\
&= (D \Join R_1) \triangleright_{p \wedge \text{ natural}} (D \Join R_2)
\end{aligned}$$

If $F(R_2) = \emptyset$, we can use a simpler transformation using Lemma 4.15:

$$\begin{aligned}
D \Join (R_1 \triangleright R_2) &= D \Join \big(R_1 \setminus \big(R_1 \ltimes_p R_2\big)\big) \\
&= (D \Join R_1) \setminus \big((D \Join R_1) \ltimes_p R_2\big) \\
&= (D \Join R_1) \triangleright_p R_2
\end{aligned}$$

□

**Lemma 4.17.** *Given a duplicate free relation $D$ and a relational expression $T := R_1 ⟕_p R_2$ with $A(D) \cap (A(R_1) \cup A(R_2)) = \emptyset$, and $F(T) \subseteq A(D)$. Then $D \Join \big(R_1 ⟕_p R_2\big) \equiv (D \Join R_1) ⟕_{p \wedge \text{ natural}} (D \Join R_2)$. For the special case that $F(R_2) = \emptyset$ we have $D \Join \big(R_1 ⟕_p R_2\big) \equiv (D \Join R_1) ⟕_p R_2$*

*Proof.* We can show this using Equation 16, Lemma 4.5, Lemma 4.9, Lemma 4.13, and Lemma 4.16:

$$\begin{aligned}
D \Join (R_1 ⟕ R_2) &= D \Join \Big(\big(R_1 \bowtie_p R_2\big) \cup \Big(\text{nullpad}_{A(R_2)}\big(R_1 \triangleright_p R_2\big)\Big)\Big) \\
&= \big((D \Join R_1) \bowtie_{p \wedge \text{ natural}} (D \Join R_2)\big) \cup \Big(\text{nullpad}_{A(D) \cup A(R_2)}\big((D \Join R_1) \triangleright_{p \wedge \text{ natural}} (D \Join R_2)\big)\Big) \\
&= (D \Join R_1) ⟕_{p \wedge \text{ natural}} (D \Join R_2)
\end{aligned}$$

If $F(R_2) = \emptyset$, we can use a simpler transformation using Lemma 4.11 and Lemma 4.16:

$$
\begin{aligned}
D \mathbin{\blacktriangleright\!\triangleleft} (R_1 \mathbin{\unicode{x27D5}} R_2) &= D \mathbin{\blacktriangleright\!\triangleleft} \Big( (R_1 \bowtie_p R_2) \cup \Big( \text{nullpad}_{A(R_2)} (R_1 \triangleright_p R_2) \Big) \Big) \\
&= \big( (D \mathbin{\blacktriangleright\!\triangleleft} R_1) \bowtie_p R_2 \big) \cup \Big( \text{nullpad}_{A(D) \cup A(R_2)} \big( (D \mathbin{\blacktriangleright\!\triangleleft} R_1) \triangleright_p R_2 \big) \Big) \\
&= (D \mathbin{\blacktriangleright\!\triangleleft} R_1) \mathbin{\unicode{x27D5}}_p R_2
\end{aligned}
$$

□

For dependent joins, we can push the dependent join with $D$ down the left side, and the unnesting algorithm will then later unnest the right hand side using Theorem 4.1. Note that it is sufficient to push $D$ down the left hand side, as the other dependent join will then make the required attributes from $D$ (if any) available on the right hand side.

**Lemma 4.18.** *Given a duplicate free relation $D$ and a relational expression $T := R_1 \mathbin{\blacktriangleright\!\triangleleft}_p R_2$ with $A(D) \cap (A(R_1) \cup A(R_2)) = \emptyset$, and $F(T) \subseteq A(D)$. Then $D \mathbin{\blacktriangleright\!\triangleleft} \big(R_1 \mathbin{\blacktriangleright\!\triangleleft}_p R_2\big) \equiv (D \mathbin{\blacktriangleright\!\triangleleft} R_1) \mathbin{\blacktriangleright\!\triangleleft}_p R_2$.*

*Proof.* We can show this using Equation 11:

$$
\begin{aligned}
m_{D \mathbin{\blacktriangleright\!\triangleleft} (R_1 \mathbin{\blacktriangleright\!\triangleleft} R_2)}(x) &= m_{D\left(x_{|A(D)}\right)} \cdot m_{\text{bind}\left(R_1, x_{|F(R_1)}\right)} \Big( x_{|A(R_1)} \Big) \cdot m_{\text{bind}\left(R_2, x_{|F(R_2)}\right)} \Big( x_{|A(R_2)} \Big) \cdot 1_{p(x)} \\
&= m_{D \mathbin{\blacktriangleright\!\triangleleft} R_1} \Big( x_{|A(D) \cup A(R_1)} \Big) \cdot m_{\text{bind}\left(R_2, x_{|F(R_2)}\right)} \Big( x_{|A(R_2)} \Big) \cdot 1_{p(x)} \\
&= m_{(D \mathbin{\blacktriangleright\!\triangleleft} R_1) \mathbin{\blacktriangleright\!\triangleleft}_p R_2}(x)
\end{aligned}
$$

□

## 5. Proving Top-Down Unnesting

Using the equivalences from Section 4, we can always push a dependent join with a duplicate set $D$ down the algebra until the right hand side no longer depends on the left hand side. By using Lemma 3.1 we can then replace the dependent join with a regular join, resulting in an unnested plan. Next we will show that the algorithm from [2] does this transformation implicitly.

The algorithm has two parts, the `simpleDJoinElimination` and the general case. We can easily see that `simpleDJoinElimination` is correct because it simply uses Lemma 4.9 and Lemma 4.8 to move correlated selections and maps above the dependent join. If that is sufficient to move all dependent operators above the dependent join, we can use Lemma 3.1 and convert the dependent join into a regular join.

If that is not sufficient to eliminate dependent joins, the full algorithm starts and initializes the `UnnestingInfo` structure, pointing it to the top-most dependent join. The notation in the pseudo code largely follows the notation in this paper. The set $D$ in `UnnestingInfo` corresponds to the set $D$ computed in Theorem 4.1. The `outerRefs` variable contains its attributes, i.e., $A(D)$. One notable difference between the pseudo code and the notation used in Section 4 is the handling of *natural* conditions. In Section 4, we rewrite, e.g., $D \mathbin{\blacktriangleright\!\triangleleft} \big(R_1 \mathbin{\unicode{x27D5}}_p R_2\big)$ into $(D \mathbin{\blacktriangleright\!\triangleleft} R_1) \mathbin{\unicode{x27D5}}_{p \wedge \text{ natural}} (D \mathbin{\blacktriangleright\!\triangleleft} R_2)$. Which means that we read $D$ on both sides and then join on the identical columns, eliminating one of the copies. The pseudo code does conceptually the same, but it introduces a new name for the columns in $D$ instead and then enforces equality of the representations. In our example expression, if $D$ consists of the column $d_1$, we would translate this into $\Big(\rho_{d_1' : d_1}(D) \mathbin{\blacktriangleright\!\triangleleft} R_1\Big) \mathbin{\unicode{x27D5}}_{p \wedge d_1' = d_1''} \Big(\rho_{d_1'' : d_1}(D) \mathbin{\blacktriangleright\!\triangleleft} R_2\Big)$. Here, $d_1'$ and $d_1''$ are the new names for $d_1$, and in the result one of the two will be picked as the new representative for the column $d_1$. (It does not matter which column is picked as both are equal). This explicit rename was chosen in pseudo code instead of the implicit natural join in algebra because the underlying database system Umbra requires all columns to have unique names.

After setting up `UnnestingInfo`, the algorithm recurses through the tree and calls `unnest` on each operator. That corresponds to the individual equivalences in Section 4; the additional `rewriteColumns` call then maps the columns from $D$ to the new representatives in the result. When encountering another dependent join, the code first unnests the left side (Lemma 4.18) and then computes a new set $D$ for the right side using Theorem 4.1, recursing as needed. For this to be sound, $A(D)$ must remain disjoint not only from the top-level schema but from every intermediate schema encountered during the recursion. If a descended-into operator would reintroduce an attribute of $A(D)$, the algorithm renames it first.

The recursion stops when the current tree no longer depends on the outer join in `UnnestingInfo`. This must always be the case at some point because the leaf nodes (usually base tables) do not depend on a dependent join. Then the code either uses Lemma 3.1 and replaces the current operator $o$ with info.$D \bowtie o$ (picking new names for the columns in $D$). Or, if we found equivalences for all columns in $D$, we can replace $o$ with $\chi_{d':\text{repr}(d)}(o)$, where $\text{repr}(d)$ is a representative for the column $d$. For example in the expression $\sigma_{d=x}(R)$, the recursion would stop at $R$ and would have collected the equivalence $d = x$. Then it could rewrite this expression into $\sigma_{d'=x}(\chi_{d':x}(R))$, picking $d'$ as new name for $d$. Note that using $\chi$ instead of $\bowtie$ computes a superset of the tuples, as the result can contain values that were not contained in $D$ (see Lemma 4.2). In Theorem 4.1 we have shown that this is safe, the final join will eliminate all values that were not included in $D$. Whether or not $\chi$ should be used instead of $\bowtie$ is a decision for the query optimizer that has to be made based upon costs.

The unnesting algorithm visits each operator at most once, starting from dependent joins in top-to-bottom order and stopping when the current tree no longer depends on the dependent joins. Afterwards, no dependent join remains in the algebra expression. Thus, the algorithm from [2] terminates and correctly unnests arbitrary queries.

## References

1. Neumann, T., Kemper, A.: Unnesting Arbitrary Queries. In: Datenbanksysteme für Business, Technologie und Web (BTW), 16. Fachtagung des GI-Fachbereichs "Datenbanken und Informationssysteme" (DBIS), 4.-6.3.2015 in Hamburg, Germany. Proceedings. pp. 383–402. GI (2015)
2. Neumann, T.: Improving Unnesting of Complex Queries. In: Datenbanksysteme für Business, Technologie und Web (BTW), 21. Fachtagung des GI-Fachbereichs "Datenbanken und Informationssysteme" (DBIS), 5.-7.3.2025 in Bamberg, Germany. Proceedings (2025)

Technische Universität München
*Email address:* neumann@in.tum.de